# Comment on the Reply to "Comment on 'Dynamically maintained steady-state pressure gradients'"

Yingbin Ge

Department of Chemistry, Central Washington University, Ellensburg, Washington 98926, United States

**Abstract:** In *Reply to "Comment on 'Dynamically maintained steady-state pressure gradients'"*, Sheehan offered his responses to the fifteen proposed resolutions to Duncan's paradox. This comment on Sheehan's reply aims to point out that it is epicatalysis that results in Duncan's paradox and the violation of the second law of thermodynamics. Until solid experimental findings unequivocally support epicatalysis, the second law of thermodynamics remains valid and Duncan's paradox is not a paradox.

In Reply to "Comment on 'Dynamically maintained steady-state pressure gradients'", Sheehan offered his responses to the fifteen proposed resolutions to Duncan's paradox.[1,2] Out of the fifteen resolutions, epicatalysis is the origin of Duncan's paradox as illustrated in Figure 1. Step 1 in Figure 1 involves epicatalysis that changes not only the kinetics but also the chemical equilibrium in an isolated system. Without losing generality, catalyst 1 (C1) selectively converts species A to B while catalyst 2 (C2) selectively converts species B to A. Step 2 in Figure 1 involves a spontaneous mixing of A and B after the catalysts are covered. Because both step 1 and step 2 are spontaneous, the system entropy should increase after these two steps. Meanwhile the entropy should remain the same as the two steps constitutes a cyclic process. The above *reductio ad absurdum* analysis rules out the possibility of epicatalysis. One may argue that the second law of thermodynamics may not hold in epicatalysis and that spontaneous changes (such as epicatalysis) in an isolated system may result in entropy decrease, but this argument is no better than the circular reasoning criticized by Sheehan in his Reply.[1]

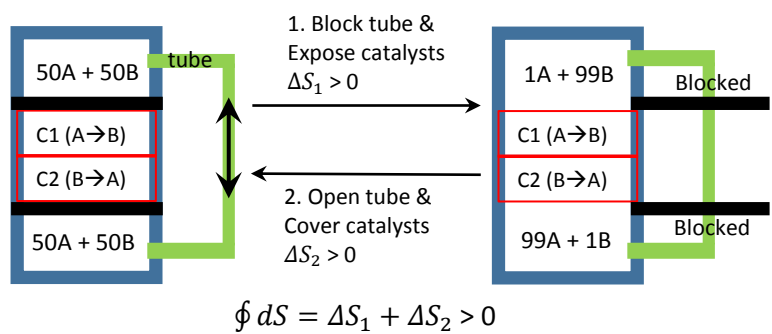

**Figure 1.** An isolated system undergoes epicatalysis followed by spontaneous mixing.

In conclusion, effort was made to challenge the second law of thermodynamics, but until one day solid experiments demonstrate that a catalyst can indeed change the chemical equilibrium, invalidating the second law of thermodynamics is out of reach. And until that day, Duncan's paradox is not a paradox.

**References**

(1) Sheehan, D. P. Reply to "Comment on 'Dynamically Maintained Steady-State Pressure Gradients.'" *Phys. Rev. E* **2000**, *61* (4), 4662–4665.

(2) Duncan, T. L. Comment on "Dynamically Maintained Steady-State Pressure Gradients." *Phys. Rev. E* **2000**, *61* (4), 4661.